\documentstyle[12pt]{article}
\begin{document}
\thispagestyle{empty}

\begin{center}
\LARGE \tt \bf{A comment on inhomogeneous Einstein-Cartan-Kalb-Ramond fields in cosmology}
\end{center}

\vspace{2.5cm}

\begin{center} {\large By  L.C. Garcia de Andrade\footnote{Departamento de
F\'{\i}sica Te\'{o}rica - Instituto de F\'{\i}sica - UERJ

Rua S\~{a}o Fco. Xavier 524, Rio de Janeiro, RJ

Maracan\~{a}, CEP:20550-003 , Brasil.}}
\end{center}

\vspace{2.0cm}

\begin{abstract}
Purely time dependent solutions in four-dimensional Einstein-Cartan-Kalb-Ramond (ECKR) theory of gravity are shown not to be possible leading to a trivial vanishing of all Kalb-Ramond fields. This result seems to contradict previously results obtained by SenGupta et al. (Class. and Quantum Gravity 19(2002)677) where de Sitter spacetime solution is found as an example of opticaly active spacetime. It seems that only inhomogeneous KR fields are possible in four-dimensional torsioned spacetimes. 
\end{abstract}

\newpage
\pagestyle{myheadings}
\markright{\underline{ECKR cosmology}}
\section{Introduction}

Kalb-Ramond fields have been used in physics on a variety of applications that run from supergravity \cite{1} to electrodynamics \cite{2} and wormholes \cite{3}. Recently SenGupta and his group put forward a series of papers \cite{2,4,5} which range from a broad spectrum of applications of Kalb-Ramond geometry playing the role of spacetime torsion \cite{2} which runs from neutrino helicity-flip untill the question of the coupling of torsion with electromagnetic fields. One of the few things that has not been handled so far is the difficult issue of time dependent solutions. History of General Relativity indeed shows that this is really true as can be understood by comparison of Schwarzschild solution with the much more mathematically involved Kerr solution. Another important motivation to study time dependent solutions is in the investigation of CMBR anisotropies where parity violation is achieved in gravito-electromagnetic fields. But an important issue on this matter is the existence of these solutions. In one of the SenGupta papers he and his group examine the question if the KR fields can make the spacetime optically active which from the physical point of view may lead to important applications in galaxy astronomy and cosmology. In this paper they made use of a de Sitter type cosmology in the investigation of inhomogeneous KR geometry. In this comment we show that actually there is no such solution and indeed there is no any other purely time dependent homogeneous solution of the ECKR field equations of gravity. The proff is straightforward and comes from the ECKR system \cite{6}
\begin{equation}
e^{-{\lambda}}(\frac{1}{r^{2}}-\frac{{\lambda}'}{r})-\frac{1}{r^{2}}=k(h_{1}h^{1}+h_{2}h^{2}+h_{3}h^{3}-h_{4}h^{4})
\label{1}
\end{equation}

\begin{equation}
e^{-{\lambda}}(\frac{1}{r^{2}}-\frac{{\nu}'}{r})-\frac{1}{r^{2}}=k(h_{1}h^{1}+h_{2}h^{2}-h_{3}h^{3}+h_{4}h^{4})
\label{2}
\end{equation}

\begin{equation}
e^{-{\lambda}}({\nu}" +\frac{{{\nu}'}^{2}}{2}-\frac{{\nu}'{\lambda}'}{2}+\frac{{\nu}'-{\lambda}'}{r})-e^{-{\nu}}(\ddot{\lambda}+\frac{{\dot{\lambda}}^{2}}{2}-\frac{\dot{\lambda}\dot{\nu}}{2})=2k(h_{1}h^{1}-h_{2}h^{2}+h_{3}h^{3}+h_{4}h^{4})
\label{3}
\end{equation}

\begin{equation}
e^{-{\lambda}}({\nu}" +\frac{{{\nu}'}^{2}}{2}-\frac{{\nu}'{\lambda}'}{2}+\frac{{\nu}'-{\lambda}'}{r})-e^{-{\nu}}(\ddot{\lambda}+\frac{{\dot{\lambda}}^{2}}{2}-\frac{\dot{\lambda}\dot{\nu}}{2})=2k(-h_{1}h^{1}+h_{2}h^{2}+h_{3}h^{3}+h_{4}h^{4})
\label{4}
\end{equation}
where we use the spherically symmetric geometry
\begin{equation}
ds^{2}=e^{{\nu}(r,t)}dt^{2}-e^{{\lambda}(r,t)}dr^{2}-r^{2}(d{\theta}^{2}+sin^{2}{\theta}{d{\phi}}^{2})
\label{5}
\end{equation}
and the remaining conditions on the ECKR gravity field are
\begin{equation}
e^{\lambda}\frac{\lambda}{r}= 2kh_{3}h^{4}
\label{6}
\end{equation}
\begin{equation}
h_{2}h^{4}=h_{1}h^{4}=h_{2}h^{3}=h_{1}h^{2}=h_{3}h^{1}=0
\label{7}
\end{equation}
Since the LHS of equations (\ref{3}) and (\ref{4}) are identical from expressions (\ref{7}) one obtains
\begin{equation}
h_{1}=h_{2}=0
\label{8}
\end{equation}
with this great simplification the KR field equations \cite{6} read
\begin{equation}
{{h^{3}},}_{3}={{h^{4}},}_{4}=0	
\label{9}
\end{equation}
which means that both KR fields are not dependent of the coordinate ${\phi}$ and finally the remaining KR field equations are
\begin{equation}
{{h^{3}},}_{2}+cot{\theta}h^{3}=0
\label{10}
\end{equation}
and
\begin{equation}
{{h^{3}},}_{0}+\frac{\dot{\nu}+\dot{\lambda}}{2}h^{3}=-{{h^{4}},}_{1}-(\frac{{\nu}'+{\lambda}'}{2} +\frac{2}{r})h^{4}
\label{11}
\end{equation}
and
\begin{equation}
{{h^{4}},}_{2}+cot{\theta}h^{4}=0
\label{12}
\end{equation}
Let us now assume by hypothesis of purely time dependent metric that the metric coefficients ${\nu}$ and ${\lambda}$ do not depend on the radial coordinate r. This is mathematically equivalent to say that their corresponding partial derivatives vanish or
\begin{equation}
{\nu}'=0={\lambda}'
\label{13}
\end{equation}
Now by applying these constraints on the equations (\ref{1}) and (\ref{2}) one obtains
\begin{equation}
e^{\lambda}=1
\label{14}
\end{equation}
or ${\lambda}=0$. This result can be immediatly used to reduce the above field equations to the following constraints 
\begin{equation}
{{h^{3}},}_{0}+(\frac{\dot{\nu}+\dot{\lambda}}{2})h^{3}=0
\label{15}
\end{equation}
which yields
\begin{equation}
h^{3}=e^{-{\frac{\nu}{2}}}
\label{16}
\end{equation}
and from equation (\ref{5})  
\begin{equation}
h_{3}h^{4}=0
\label{17}
\end{equation}
while from the RHS of equations (\ref{1}) and (\ref{2})
\begin{equation}
h_{3}h^{3}=h_{4}h^{4}=0
\label{18}
\end{equation}
These last two constraints on the solution yields that both KR fields vanish identically proving that there are no KR fields in purely time dependent spacetimes as we meant to prove. Of course general inhomogeneous cosmologies such as open or closed FRW models can be obtained.

\section*{Acknowledgement}
I am very much indebt to Prof. P.S.Letelier for helpful discussions on the subject of this  paper. Financial support from CNPq. and UERJ are gratefully ackowledged.

\end{document}